\documentclass{PoS}
\usepackage{amsmath}
\usepackage{wrapfig}
\usepackage[dvipdfmx]{}

\title{Exploring finite density QCD phase transition with canonical approach ---Power of multiple precision computation---}

\ShortTitle{Exploring finite densty QCD phase transition}

\author{\speaker{Shotaro Oka}\\
        Institute of Theoretical Physics, Department of Physics,
        Rikkyo University, Toshima-ku, Tokyo, 171-8501 Japan\\
        E-mail: \email{okasho-hato@rikkyo.ac.jp}}

\author{for Zn--Collaboration}

\abstract{The canonical approach for finite density lattice QCD has
	a numerical instability. This instability makes it difficult to use the method reliably
	at the finite real chemical potential region.
	We studied this instability in detail and found that it is caused by the cancellation
	of significant digits. In order to reduce the effect of this cancellation,
	we adopt the multiple precision calculation for our discrete Fourier
	transformation (DFT) program, and we get the canonical partition function
	$Z_\textrm{C} (n, T)$ with required accuracy.
	From the obtained $Z_\textrm{C} (n, T)$, we calculate Lee--Yang zero
	distribution varying the number of significant digits.
	As a result, some curves surround the origin in the fugacity plane,
	but they are moved by varying the number of significant digits.
	Hence, we conclude that these curves are pseudo phase transition lines,
	and not real ones.}

\FullConference{The 33rd International Symposium on Lattice Field Theory\\
		14 -18 July 2015\\
		Kobe International Conference Center, Kobe, Japan*}

\begin{document}

$\phantom{.}$\vspace{-9ex}
\section{Introduction}
For a long time, the QCD phase diagram has been studied using a perturbation theory,
some models, and heavy ion collider experiments.
However, in these methods, we can explore only a limited range of temperature and density.
Therefore, we have not exactly known the QCD phase diagram yet.

A numerical simulation of the lattice QCD is a powerful tool to study the QCD phase diagram
because this may explore finite temperature and finite density regions.
But, in the finite density region, a famous problem ---the sign problem--- occurs.
% Hence, we cannot calculate physical quantities well in this region.
Hence, we cannot calculate physical quantities in a satisfactory manner.

In this paper, we study the canonical approach\cite{canonical}.
This method may drive away the sign problem;
however, a numerical instability alternatively arises in a finite density simulation.
We will denote that we can reduce this instability,
which is necessary to calculate thermodynamic quantities
at the finite real chemical potential (finite density) region.

\section{Sign problem}
In order to calculate thermodynamic quantities,
let us consider the grand canonical partition function,
\begin{equation}
	Z_\textrm{GC} (\mu, T) = \int D U  {\left[ \det \Delta (x, x^\prime, \mu) \right]}^{N_f} e^{-S_G}, \label{grand}
\end{equation}
where $\Delta (x, x^\prime, \mu)$ is the fermion matrix, $S_G$ is gauge action,
$\mu$ is chemical potential, and $N_f$ is the number of flavor.
In this study, we use $O (a)$ improved Wilson fermions\cite{clover} and the Iwasaki gauge action\cite{Iwasaki}.

By the definition of Wilson fermions,
we can derive ${\left[ \det \Delta (\mu) \right]}^\star = \det \Delta (-\mu^\star)$.
This equation means that,
if chemical potential is zero ($\mu = 0$) or pure imaginary ($\mu = i \mu_I$),
$\det \Delta (\mu)$ is real,
and otherwise $\det \Delta (\mu)$ is complex. 
Thus, in a finite density simulation when chemical potential is real,
${\left[ \det \Delta (\mu) \right]}^{N_f}$ in Eq.(\ref{grand}) becomes complex,
and therefore the standard Monte--Carlo method does not work.
This problem is called the sign problem.

\section{Canonical approach and its numerical instability}
It is well known that the grand canonical partition function can be represented
with the canonical partition function $Z_\textrm{C} (n, T)$ as the following equation,
\begin{equation}
	Z_\textrm{GC} (\mu, T) = \sum_{n=-\infty}^{\infty} Z_\textrm{C} (n,T) {\left( e^{\mu/T} \right)}^n, \label{fug}
\end{equation}
where $n$ is net quark number.
Then, because $\exp (\mu/T)$ is fugacity, this relation is called the fugacity expansion.
We can use the fugacity expansion when we calculate the grand partition function
for arbitrary chemical potential.
Such a method ---to calculate grand canonical quantities from canonical quantities---
is called the canonical approach.

The canonical approach can drive the sign problem away from the front.
This is because, when we calculate the canonical partition function, 
we can use the following equation, \enlargethispage{0.5ex}
\begin{equation}
	Z_\textrm{C} (n, T) = \int_{0}^{2 \pi} d \left( \frac{\mu_I}{T} \right) \, Z_\textrm{GC} (\mu = i \mu_I, T) e^{-i (\mu_I/T) n}. \label{Four}
\end{equation}\newpage

$\phantom{.}$\vspace{-5ex}\\
This equation is just the Fourier transformation of $Z_\textrm{GC} (\mu, T)$
for pure imaginary chemical potential $\mu = i \mu_I$
\footnote{%Apparently
Eqs.(\ref{fug}) and (\ref{Four}) look like tautology, 
but Eq.(\ref{fug}) holds for any number $\mu$
and Eq.(\ref{Four}) holds only for pure imaginary $\mu = i \mu_I$.
They are not equivalent at all.
}.
Note that, in a numerical simulation, we use the discrete Fourier transformation (DFT) as
\vspace{-1.2ex}
\begin{equation}
	Z_\textrm{C} (n, T) = \frac{1}{N} \sum_{k=0}^{N-1} \, Z_\textrm{GC} \left( i \frac{\mu_I}{T} = i \frac{2 \pi k}{N} \right) e^{-i (2 \pi k / N) n}. \label{dFour}
\end{equation}
For pure imaginary chemical potential,
the fermion determinant $\det \Delta (\mu = i \mu_I)$ becomes real,
and no sign problem occurs
if the number of fermions $N_f$ is even.
Therefore, we may calculate $Z_\textrm{C} (n)$ using Eq. (\ref{Four}) without the sign problem.

\begin{wrapfigure}{r}{0.55\hsize}
	\centering
 	\includegraphics[width=\hsize ,clip]{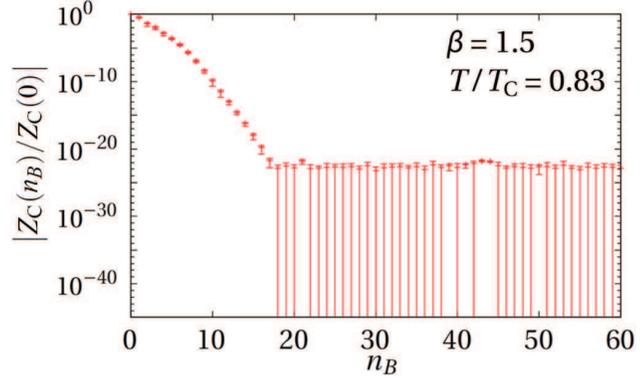}
	\caption{Normalized canonical partition function $\bigl| Z_\textrm{C} (n_B) / Z_\textrm{C} (0) \bigr|$ as a function of baryon number $n_B$ at $\beta = 1.5$ (below $T_c$).}
	\label{Zn_sat}
\end{wrapfigure}

However, it is known that the canonical approach has a numerical instability\cite{WNE}.
This instability occurs when we perform the discrete Fourier transformation in Eq.(\ref{dFour}).
Fig.\ref{Zn_sat} shows that 
this instability arises when net baryon number $n_B$ becomes large.
We can calculate the canonical partition function $Z_\textrm{C} (n)$ with enough accuracy
when $n_B$ is small (up to $n_B \simeq 15$), but, for larger $n$, we can not control them well.
This phenomenon is called a numerical instability of the canonical approach.
Because this instability arises when we calculate $Z_\textrm{C} (n)$,
the canonical approach does not work actually.\vspace{0.4ex}

On the other hand, we can use the reduction formula for the fermion determinant
when we calculate $Z_\textrm{C} (n)$\cite{deForcrand, Kentucky, NagataNakamura}.
Using this method, we can calculate $Z_\textrm{C} (n)$ exactly
because it solve a characteristic equation of the determinant
and then its eigenvalues relate to $Z_\textrm{C} (n)$.

However, this method requires large numerical cost
and its result often overflows.
The origin of this overflow is that
the magnitude of $Z_\textrm{C} (n)$ varies drastically when changing $n$.
Double precision floating point variables can represent real numbers from $10^{-308}$ to $10^{308}$,
but the magnitude of $Z_\textrm{C} (n)$ varies beyond this range.
We can reduce this overflow by increasing the range of number.
Specifically, we can use wide--range numbers\cite{XQCD-J}
or the multiple precision calculation\cite{deForcrand, Kentucky};
however, the numerical cost becomes even larger.

In this paper, we will study how to get $Z_\textrm{C} (n)$ with enough accuracy from a different viewpoint.

\vspace{-0.4ex}
\section{Solution of numerical instability}
Since Eq.(\ref{Four}) can be proved analytically,
we can consider that this numerical instability is caused by numerical errors.
Numerical errors can be classified into following types of error (see also Refs.\cite{TheArt} and \cite{guardbit}).
\begin{enumerate}
	\setlength{\parskip}{0ex} % 段落間
	\setlength{\itemsep}{0.4ex} % 項目間
	\renewcommand{\labelenumi}{(\arabic{enumi})}
	\item Rounding error: 
	for a numerical calculation, numerical values are always rounding when they are assigned to each variable
	% since variables have only the finite number of significant digits.
	since variables have only the finite precision.
	Hence, there is a difference between a numerical value and its exact mathematical value;
	this difference is called the rounding error.
	For example, this error arises in the following type of calculation:
	
	$\phantom{.}$\vspace{-5ex}
	\begin{equation}
		\begin{array}{lll} \displaystyle
			\frac{1}{3} = & 3.333333\cdots \times 10^{-1} \xrightarrow{\textrm{assignment to a variable}} & 3.333333 \times 10^{-1}. \\
	% 		& \textrm{(Exact mathematical representation)} & \textrm{(Numerical representation)}
			& \textrm{(Exact mathematical value)} & \textrm{(Numerical value; 7 digits)}
		\end{array} \label{round_err}
	\end{equation}
	In above case, the exact mathematical value has infinite significant digits;
	however the numerical value has only seven significant digits.
	The numerical result is less than the exact mathematical one as $3.333\cdots \times 10^{-8}$.
	Then, to perform rounding exactly, computers usually have a guard digit\cite{guardbit}.
% 	For instance, the standard of floating point number operation, IEEE 754,
% 	% requires one guard bit for a floating point number.
% 	requires three bits (they are called the guard bit, round bit, and sticky bit, respectively) for a floating point number operation.
% 	% \footnote{
% 	% 	IEEE 754 also requires a round bit and a sticky bit.
% 	% 	These bits is used when rounding processing with guard bit.
% 	% }.
% 	When rounding the exact value in Eq.(\ref{round_err}),
% 	a computer puts first eight significant digits on a hardware, 
% 	and it rounds eight--th significant digit,
% 	finally it assigns rounded seven significant digits to a variable on a memory.
% 	Such eight--th significant digit corresponds to guard bits.

	\item Truncation error:
	when we estimate an infinite series numerically,
	we must truncate it.
	Then, there is a difference between the exact mathematical result and its truncated one.
	This difference is called the truncation error.
	We can estimate the effect of this error by varying a truncation point of a series.

	\item Cancellation of significant digits:
	% when we calculate a subtraction between two numerical values which are almost the same,
	when we calculate a subtraction between two nearly equal values,
	a lot of higher significant digits are cancelled,
	therefore the result has only a few significant digits
	since the number of significant digits is limited for numerical calculation.
	% Because this, the rounding error is transmitted to higher significant digits from lowest significant digit.
	This phenomenon is called the cancellation of significant digits.
	Specifically, it occurs the following type of calculation:
	\begin{equation}
		\begin{array}{lll}
			1.234567 - 1.234566 & = & 0.000001. \\
			\textrm{(7 significant digits)} & & \textrm{(1 significant digit)}
		\end{array} \label{drop_1}
	\end{equation}
	In this case, variables have seven significant digits, and six significant digits are lost in this subtraction.
	If we want to reduce the effect of this cancellation,
	we should increase the number of significant digits.
	For instance, we can consider the following calculation
	instead of above one:
	\begin{equation}
		\hspace{-0.3em}
		\begin{array}{lll}
			1.234567444444444444444 - 1.234566111111111111111 & = & 0.000001333333333333333. \\
			\textrm{(22 significant digits)} &  & \textrm{(16 significant digits)}
		\end{array} \label{drop_16}
	\end{equation}
	Six significant digits are similarly lost in this calculation,
	and yet 16 significant digits still remain in the final result.

	\item Loss of trailing digits:
	when we calculate an addition or subtraction between a huge number and small one,
	many lower significant digits of the small number are cut since variables hold only the finite precision.
	Then, the numerical result is underestimated than the mathematical result.
	This phenomenon is the loss of trailing digits.
	Specifically, it arises the following type of calculation:
	\begin{equation}
		\begin{array}{lll}
			3 \times 10^{10} + 2 \times 10^{0} = & 3.000000002 \times 10^{10} \xrightarrow{\textrm{assignment}} & 3.000000 \times 10^{10}. \\
			&\textrm{(Exact mathematical result)} & \textrm{(Numerical result; 7 digits)}
		\end{array} \label{loss_7}
	\end{equation}
	In this case, variables have seven significant digits,
	and the numerical result is underestimated than the mathematical one as $2 \times 10^{-1}$.
	% Lower seven significant digits thus are lost.
	As the cancellation of significant digits, 
	we should increase the number of significant digits if we want to reduce this loss.
	\begin{equation}
		\begin{array}{lll}
			3 \times 10^{10} + 2 \times 10^{0} = & 3.000000002 \times 10^{10} \xrightarrow{\textrm{assignment}} & 3.000000002000000 \times 10^{10}. \\
			&\textrm{(Exact mathematical result)} & \textrm{(Numerical result; 16 digits)}
		\end{array} \label{loss_10}
	\end{equation}
	In this calculation, by increasing significant digits,
	we can reduce an error between the mathematical result and numerical one.
\end{enumerate}
	
$\phantom{.}$\vspace{-5ex}\\
In order to make the method to be reliable and promising tool,
it is important to realize what occurs in the canonical approach.

Let us consider each type of error.
Among these, we can neglect the rounding and the truncation error
because of following reasons.
\begin{enumerate}
	\setlength{\parskip}{0ex} % 段落間
	\setlength{\itemsep}{0.4ex} % 項目間
	\renewcommand{\labelenumi}{(\arabic{enumi})}
	\item Usually, the rounding error is not transmitted to higher significant digits
	since computers have a guard digit\footnotemark.
	% This error can be transmitted only through the cancellation of significant digits
	% and the loss of trailing digits.
	If this error seems to be transmitted to higher significant digits,
	then the cancellation of significant digits or the loss of trailing digits take place.
	In this paper, we consider that the transmission of rounding error is just an effect of
	the cancellation and loss.
	% On the other hand, the rounding error is caused by some types of rounding
	% (rounding off fractions, rounding up fractions, rounding to nearest, etc.),
	% but the cancellation and loss are essentially caused only by the rounding down fractions.
	% Therefore, we discuss them as different types of error.

	\item We do not truncate Fourier series in Eq.(\ref{dFour}).
	However, if we use the hopping parameter expansion or the winding number expansion
	when we calculate the fermion determinant $\Delta (\mu)$,
	we should truncate an infinite series of these expansion.
	Then, a truncation error arises,
	but we can control it by changing a truncation point of this series.
\end{enumerate}
\footnotetext{
	See Ref.\cite{guardbit} for the guard bit;
	it is a detailed review for modern floating point arithmetic.
}
\vspace{-0.4ex}
Therefore, this numerical instability is caused by
the cancellation of significant digits, the loss of trailing digits, or both.\vspace{0.4em}

In this work, we actually monitored the behavior of all variables in our
DFT program to study the effect of these two errors.
% (Our numerical setup will be described later.)
As a result, we found that this numerical instability is caused by
the cancellation of significant digits.
This cancellation arises mainly at the last addition of the summation
in Eq.(\ref{dFour}); specifically,
% \vspace{-0.2ex}
\begin{align}
	Z_\textrm{C} (n, T) &= \frac{1}{N} \left[ \left( \sum_{k=0}^{N-2} \, Z_\textrm{GC} \left( \frac{2 \pi k}{N} \right) e^{-i (2 \pi k / N) n} \right) + Z_\textrm{GC} \left( \frac{2 \pi (N-1)}{N} \right) e^{-i (2 \pi (N-1) / N) n} \right].  \label{dFour_plus} \vspace{-0.2ex} \\
		&\hspace{15.3em} \ \uparrow \textrm{This addition.} \notag
\end{align}
\vspace{-5ex}
%\enlargethispage{0.4ex}
\begin{wrapfigure}{r}{0.55\hsize}
	\centering
 	\includegraphics[width=\hsize ,clip]{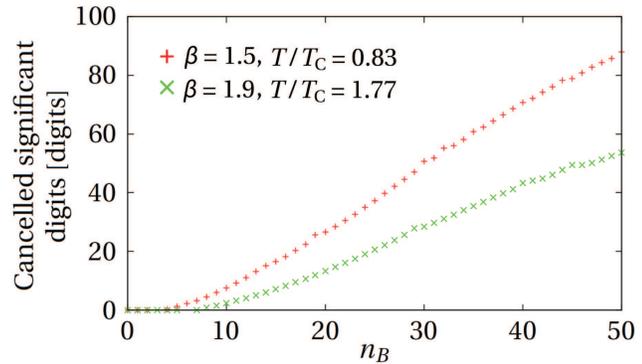}
	\caption{Cancelled significant digits of DFT program as a function of net baryon number $n_B$.
		Upper red points are data below $T_c$ and lower green points are data above $T_c$.}
	\label{FFT_drop}
\end{wrapfigure}
Fig.\ref{FFT_drop} shows how many digits are lost at this addition
in Eq.(\ref{dFour_plus}).
For example, at $n_B = 50$, about 90 significant digits are lost below $T_c$
and about 50 digits are lost above $T_c$.
From this figure, we can see that, in both temperature,
this cancellation becomes serious when $n_B$ becomes large.

As explained above, the cancellation of significant digits can be
reduced by increasing the number of significant digits.
\begin{figure}
	\vspace{-3ex}
	\centering
 	\includegraphics[width=\hsize ,clip]{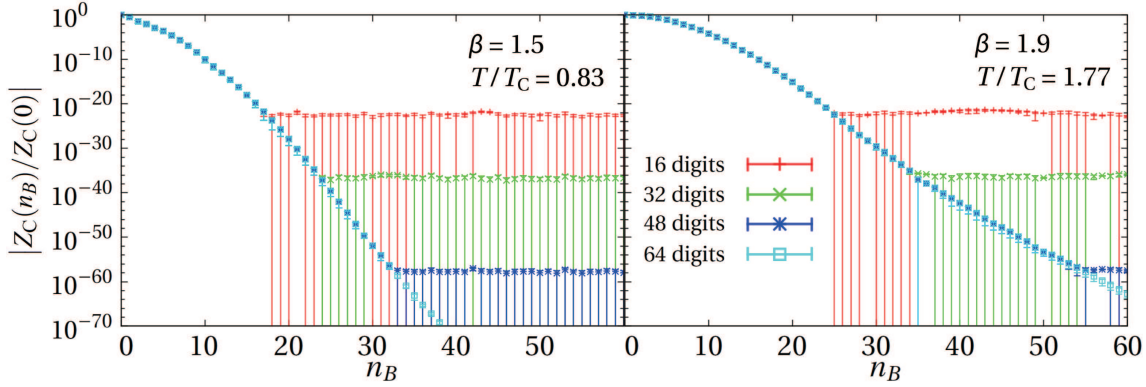}
	\caption{Normalized canonical partition function as a function of $n_B$
		and the number of significant digits. Left panel shows data below $T_c$ and right
		panel shows data above $T_c$. In both temperature, red, green, blue, and cyan points
		are calculated with 16, 32, 48, and 64 significant digits, respectively.}
	\label{Zn_drop}
\end{figure}
Fig.\ref{Zn_drop} shows the relation between a result of $Z_\textrm{C} (n)$
and the number of significant digits.
From this figure, we can see that, by increasing the number of significant digits,
we can get $Z_\textrm{C} (n)$ very reliably in both temperature.
% Increasing significant digits plays an important role in a calculation of $Z_\textrm{C} (n)$.\vspace{0.4em}

$Z_\textrm{C} (n)$ becomes small rapidly when increasing net quark number $n$.
Then, the cancellation of significant digits becomes severe.
Since $Z_\textrm{C} (n)$ is just a residue of this cancellation,
the number of \textit{cancelled} significant digits in Fig.\ref{FFT_drop} directly relates to
a magnitude of $Z_\textrm{C} (n)$ in Fig.\ref{Zn_drop}.\vspace{0.4em}

When we change the lattice volume $V$ with fixing the number of significant digits,
a saturation point (kink) of $Z_\textrm{C} (n)$ moves.
This is because the net quark number density $n / V$ varies when we change $V$ and fix $n$.
When $n / V$ is increased and temperature $T$ is fixed,
$Z_\textrm{C} (n, V, T)$ becomes small since it then corresponds to an occurrence probability of excited states. 
If we increase $n$ and fix $V$ and $T$, $Z_\textrm{C} (n, V, T)$ becomes small as Fig.\ref{Zn_drop}.
However, if we increase $V$ and fix $n$ and $T$, $Z_\textrm{C} (n, V, T)$ becomes large.
This means that a kink of $Z_\textrm{C} (n, V, T)$ moves toward larger $n$ region when we increase lattice volume $V$.
\vspace{-1.2ex}

\section{Lee--Yang zero distribution using multiple precision canonical approach}

% $\phantom{.}$\vspace{-0.4ex}
\begin{wraptable}{r}{0.55\hsize}
	\vspace{-5ex}
	\centering
		\begin{tabular}{cccccc}
			$\beta$ & $C_\textrm{SW}$ & $\kappa$ & $T / T_c$ & $m_\pi / m_\rho$ & \#Conf. \\ \hline
			$1.5$ & $1.1$ & $0.136$ & $0.83$ & $0.756(13)$ & $100$ \\
			$1.9$ & $1.1$ & $0.125$ & $1.77$ & $0.714(15)$ & $600$
		\end{tabular}
	\caption{Parameters of each simulation.}
	\label{param_table}
\end{wraptable}
\vspace{-0.4ex}

\subsection{Our numerical setup}

We use the Iwasaki gauge action\cite{Iwasaki} and
clover improved Wilson fermions\cite{clover}.
When we calculate $\det \Delta (\mu)$ at pure imaginary chemical potential,
we use the winding parameter expansion\cite{WNE} and a reweighting technique
(see Ref.\cite{Nakamura_talk}).
The number of flavor is set to $N_f = 2$.

We use the APE stout smeared gauge link\cite{APE}.
Tab.\ref{param_table} shows parameters of each simulation.
The multiple precision calculation is applied only in our DFT program and
Lee--Yang zero exploring program,
where the number of significant digits is set to 400.
In other words, we make gauge configurations and calculate $W_m$
with double precision variables (16 significant digits).
In our DFT program, the division number of an integration interval is $N = 512$.

\subsection{Results of Lee--Yang zero distribution}\enlargethispage{1ex}

Lee--Yang zeros (LYZ)\cite{YangLee} are zeros of the grand partition function on the complex chemical potential
or fugacity plane.
These points represent phase transition points,
and we can thus extract some informations of phase transition from their distribution.

\begin{figure}
	\vspace{-3ex}
	\centering
 	\includegraphics[width=0.72\hsize ,clip]{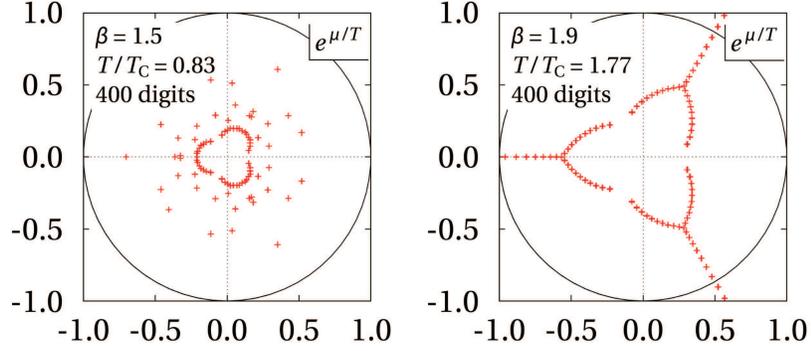}
	\caption{LYZ distribution of $8^3 \times 4$ lattice on the fugacity $\exp (\mu/T)$ plane. Left panel shows data below $T_c$ and right panel shows data above $T_c$. Here, black circles are just unit circles.}
	\label{LYZ_8884}
	\vspace{0.4ex}
\end{figure}
Fig.\ref{LYZ_8884} shows LYZ distribution below $T_c$ and above $T_c$
(lattice size: $8^3 \times 4$).
Above $T_c$, three straight lines extend radially to origin.
Cross points of these lines and the unit circle correspond
to the Roberge--Weiss phase transition\cite{RW} that is a characteristic phase transition
in the deconfinement phase at pure imaginary chemical potential.
In both temperature, there are some curves which are surrounding
the origin. We might usually think that these curves are phase transition lines
and the inside and outside of these curves are different phase.
However, as described below, it is an artifact.

\vspace{0.8ex}
\begin{figure}
	\centering
 	\includegraphics[width=\hsize ,clip]{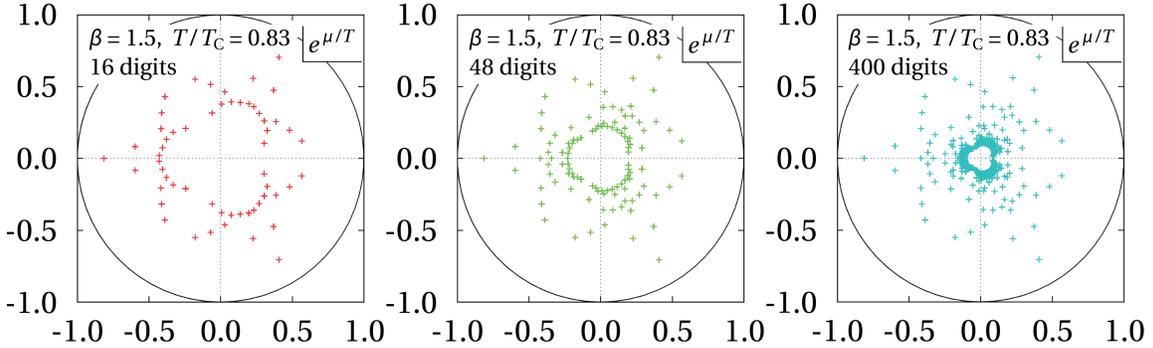}
	\caption{LYZ distribution of $12^3 \times 4$ lattice below $T_c$
		on the fugacity $\exp (\mu/T)$ plane.
		Red, green, and cyan points are calculated with 16, 48, and 400 significant digits, respectively.
		Here, black circles are just unit circles.}
	\label{LYZ_12121204}
\end{figure}
\vspace{0.8ex}
Fig.\ref{LYZ_12121204} shows LYZ distribution below $T_c$
when varying the number of significant digits
\footnote{
	Then, we check $Z_\textrm{C} (n)$ results which are calculated with 16, 48, and 400 significant digits,
	and if the saturation (kink) of $Z_\textrm{C} (n)$ appears,
	we then truncate each $Z_\textrm{C} (n)$ at this point, respectively.
	After these processing, we calculate Lee--Yang zeros using Eq.(\ref{fug})
	with 16, 48, and 400 digits.
} (lattice size: $12^3 \times 4$; this is a preliminary result).
Such curves are moving when we vary the number of significant digits.
Thus, we conclude that these curves are pseudo phase transition lines,
and not real ones.
\vspace{-1.2ex}

\acknowledgments
\vspace{-2ex}
This study was conducted for Zn--Collaboration.
The author is grateful to members of Zn--Collaboration,
R. Fukuda, A. Nakamura, S. Sakai, A. Suzuki, and Y. Taniguchi for their powerful support.
The author thanks Prof. T. Eguchi for his valuable discussion and encouragement.
The author appreciates very useful comments by Prof. Ph. de Forcrand and Prof. A. Alexandru.
This work is in part based on Lattice QCD common code Bridge++.
These calculations were performed using SX--9 and SX--ACE at Research Center
for Nuclear Physics and SR16000 at Yukawa Institute
for Theoretical Physics.
This work is partially supported by Kakenhi 15H03663 and 26610072.

\end{document}